\documentclass[manuscript]{acmart}
\renewcommand\footnotetextcopyrightpermission[1]{} 
\usepackage{graphicx}
\AtBeginDocument{%
  \providecommand\BibTeX{{%
    \normalfont B\kern-0.5em{\scshape i\kern-0.25em b}\kern-0.8em\TeX}}}

\setcopyright{acmcopyright}
\copyrightyear{2018}
\acmYear{2018}
\acmDOI{XXXXXXX.XXXXXXX}

 \acmBooktitle{} 
\acmPrice{15.00}
\acmISBN{978-1-4503-XXXX-X/18/06}

\begin{document}

\title{How many others have shared this? Experimentally investigating the effects of social cues on engagement, misinformation, and unpredictability on social media}

\author{Ziv Epstein}
\authornote{Both authors contributed equally to this research.}
\email{zive@mit.edu}
\affiliation{%
  \institution{MIT Media Lab}
  \streetaddress{75 Amherst St}
  \city{Cambridge}
  \state{MA}
  \country{USA}
  \postcode{02139}
}

\author{Hause Lin}
\authornotemark[1]
\affiliation{%
  \institution{University of Regina}
  \country{Canada}
}
\affiliation{%
  \institution{Sloan School of Management}
  \streetaddress{75 Amherst St}
  \city{Cambridge}
  \state{MA}
  \country{USA}
  \postcode{02139}
}
\email{hause@mit.edu}

\author{Gordon Pennycook}
\affiliation{%
  \institution{University of Regina}
  \city{Regina}
  \country{Canada}
}

\author{David Rand}
\affiliation{%
 \institution{Sloan School of Management, MIT}
 \country{USA}}

\renewcommand{\shortauthors}{Epstein and Lin, et al.}

\begin{abstract}
 Unlike traditional media, social media typically provides quantified metrics of how many users have engaged with each piece of content. Some have argued that the presence of these cues promotes the spread of misinformation. Here we investigate the causal effect of social cues on users' engagement with social media posts. We conducted an experiment with N=628 Americans on a custom-built newsfeed interface where we systematically varied the presence and strength of social cues. We find that when cues are shown, indicating that a larger number of others have engaged with a post, users were more likely to share and like that post. Furthermore, relative to a control without social cues, the presence of social cues increased the sharing of true relative to false news. The presence of social cues also makes it more difficult to precisely predict how popular any given post would be. Together, our results suggest that -- instead of distracting users or causing them to share low-quality news -- social cues may, in certain circumstances, actually boost truth discernment and reduce the sharing of misinformation. Our work suggests that social cues play important roles in shaping users' attention and engagement on social media, and platforms should understand the effects of different cues before making changes to what cues are displayed and how.
 \end{abstract}

\ccsdesc[500]{Human-centered computing~Empirical studies in HCI}

\keywords{social influence, misinformation, social media, platform design}

\maketitle

\section{Introduction}
In recent years, social media has emerged as an important medium for citizens to consume news, with nearly half of Americans getting news from social media in 2021 \cite{walker2021news}. A unique feature of social media relative to other forms of media such as radio, TV or print is that social media platforms display a large volume of highly quantified social information about others' behavior. In particular, posts a user sees on social media are often annotated with quantified social cues, indicating how many other people have engaged with that post. 

A growing body of work has explored the role of potential social influence cues on information environments outside the context of news \cite{salganik2006experimental, epstein2021social, abeliuk2017taming, hogg2014disentangling}. This literature focuses around the idea that social influence leads to winner-take-all market characterized by high degrees of popularity inequality and unpredictability. However, these findings have yet to be explored in the context of a social media feed, with posts of a varying degree of quality. 

In parallel, a body of descriptive work has explored the association of social cues with news sharing \cite{chen2021makes, brady2017emotion, berger2021makes}, particularly in the context of the spread of misinformation on social media. Yet to date there is very little experimental investigation of the \textit{causal} effect of social engagement cues on specifically how users engage with content on social media. To address this issue, we set out with the following research questions:

\begin{enumerate}
    \item Does seeing that a larger number of others have engaged with a post increase the likelihood that users share or like it themselves?
    \item Does the presence of social cues make users less sensitive to the veracity of the news they choose to engage with?
  \item Does social influence decrease the predictability of whether posts will be successful (i.e. more likely to be engaged with)?
\end{enumerate}

In this paper, we present an experiment in which participants scrolled through a social media feed in a custom-developed web application designed to emulate the social media environment. Participants were randomly assigned to either a control condition, where posts had no social cues whatsoever, or a treatment condition, where each post was annotated with social cues indicating a (randomized) number of likes and shares. By randomizing the number of shares and likes that are displayed, we can isolate the causal effect of that information in and of itself, separate from features of the content that would, in reality, lead to higher or lower rates of engagement.

Within the treatment condition, we find that social cues indicating higher rates of engagement cause participants to be more likely to engage with the posts themselves -- and that this effect occurs to the same extent for true versus false news. This finding implies a ``rich gets richer'' effect (or "Matthew effect") which drives inequality of success, building on \citet{salganik2006experimental}. Furthermore, showing social cues in the treatment increases sharing discernment (i.e. the quality of news shared) relative to the control (where no cues were displayed). Finally, we find that social cues decrease the predictability of post sharing, which is also in line with \citet{salganik2006experimental}. 

Taken together, these results highlight what is at stake when designing social cues into social media. By increasing the likelihood a user will engage with a post, these cues increase the inequality of popularity, as well as the unpredictability of success. However, we also found that social cues improved the quality of the information ecosystem: contrary to previous work that suggests that social cues drive the spread of misinformation \cite{avram2020exposure}, our results suggest that social cues can actually boost sharing discernment in certain circumstances.

\section{Related Work}
Previous work has explored why people share content on social media. \citet{chen2021makes} find that sharing is positively predicted by a combination of a post’s perceived accuracy and familiarity, as well as its perceived importance and emotional evocativeness. The authors also find heterogeneity, such that participants with higher cognitive reflection and political knowledge, and that those who are more politically liberal put less weight on the importance/emotional evocativeness factor. \citet{berger2012makes} find that positive valence as well as high-arousal content are more likely to go viral online than negative valence and low-arousal content. Related concepts have also been associated with content diffusion, such as how interesting content is \cite{bakshy2011everyone}, how funny it is \cite{warren2011influence}, how much sentiment volatility \cite{berger2021makes} and moral-emotional language \cite{brady2017emotion} it contains, and how useful the information is \cite{heath2001emotional}.

There have also been attempts to explore how elements of social media may impact perceptions of content credibility and therefore the spread of misinformation. \citet{hameleers2020picture} find that multi-modal (i.e., headline text and image composed together) misinformation is considered more credible than the corresponding textual content itself. \citet{shen2019fake} find individual differences such as internet skills, photo-editing experience, and social media usage are associated with higher accuracy in image credibility evaluation.  \citet{mitra2017parsimonious} find several linguistic classes of words that are predictive of credibility, such as subjective words relating indicating perfection (e.g. immaculate), agreement (e.g. unanimous), newness (e.g. unique), awe and wonder (e.g. vibrant), positive and negative emotionally charged words (e.g. sucky and eager), hedges (e.g. appeared) and boosters (e.g. undeniably), evidentials (e.g. tell), anxiety (e.g. distress) and conjunctions (e.g. while). These findings are important because understanding the design features that allow users to discern the credibility of social media posts may help inspire the next generation of interfaces. 


In parallel, a growing body of work in HCI has explored lightweight design interventions as an approach to reduce misinformation sharing. \citet{jahanbakhsh2021exploring} find that providing an accuracy assessment and corresponding rationale at post time can reduce the sharing of false content.  \citet{yaqub2020effects} explore the effectiveness of credibility indicators (e.g. warning labels) for mitigating false news sharing, and find that the effectiveness of the indicators varied, with those attributed to fact checking being the most effective, and those attributed to AI being the least effective. \citet{epstein2022explanations} find that providing explanations to credibility indicators attributed to AI can increase their effectiveness. 

Focusing on social cues in particular, a growing body of work has looked at the role of such social cues in online engagement, from settings such as online advertisement \cite{bakshy2012social, huang2020social} and voting behavior\cite{jones2017social}. At the population level, \citet{salganik2006experimental} find that social influence leads to inequality and unpredictability (e.g. variance in success across worlds) of content success: it makes more popular content more popular and unpopular content less popular, but it is difficult to precisely predict the popularity of the content. In a conceptual replication, \citet{epstein2021social} find that in the context of cultural artifacts (where popularity is defined subjectively), the unpredictability that social influence induces corresponds to diverse local trends that diverge from the status quo. In other words, social influence can jumpstart niche aesthetic trends by demonstrating social proof of the concept. However, in ``objective'' contexts where content has some intrinsic quality such as veracity, social influence can actually undermine the wisdom of the crowd by diminishing the crowd's diversity \cite{lorenz2011social}.

Perhaps most relevant to our work is \citet{avram2020exposure}, who explored the impact of such social cues on social media engagement in an online experiment with over 8,500 participants. They randomly generated a social media cues for each presented article and found that social cues can strongly influence engagement with low quality information (e.g. posts from low-credibility domains, sourced from the Hoaxy API). However, their study has several limitations. For one, players had the option to fact-check articles, a feature that deviates from the actual experience of social media, and which may have inadvertently prompted people to think more directly about accuracy than they otherwise might \cite{pennycook2021shifting}. Indeed, past studies have shown that simply asking about accuracy can decrease the sharing of content that people are able to identify as being false \cite{pennycook2021shifting, pennycook2020fighting, pennycook2022accuracy}. Perhaps most importantly, the authors do not include a control condition without any social cues, which prevents causal inferences about the role of social cues relative to a baseline. 

\section{Methods}
We recruited a convenience sample of N = 644 Americans, of which N = 628 completed the survey, using the recruitment platform Prolific. Our sample had mean age = 35.7, 46.5\% female, and 66\% white. 

Participants were routed to Yourfeed, a website we designed that displays content in a feed layout. The user interface of Yourfeed mirrors the appearance of commonly used social media sites, such as Facebook or Twitter. At the onset of the experience, participants saw a modal that said ``Thank you for participating! Next, you will see a social media newsfeed, configured just for you. Please browse this newsfeed like you usually would for social media. For each post, indicate whether you would consider sharing it with your network.''

Yourfeed displayed 120 social media posts to each user in their feed, and users could click to like or share any posts they choose. Of the 120 posts shown, 90 were randomly sampled from a set of 200 political and non-political news items \cite{pennycook2021practical}, half of which are true and half false. The other 30 were randomly sampled from a set of 76 opinion and mundane news items. The mundane posts were sourced from tabloid sites (e.g. The Sun, Metro, Daily Mail) and opinion posts are opinion pieces from reputable sources such as the New York Times and the Economist. 

\begin{figure}[h]
    \centering
    \includegraphics[width=0.99\textwidth]{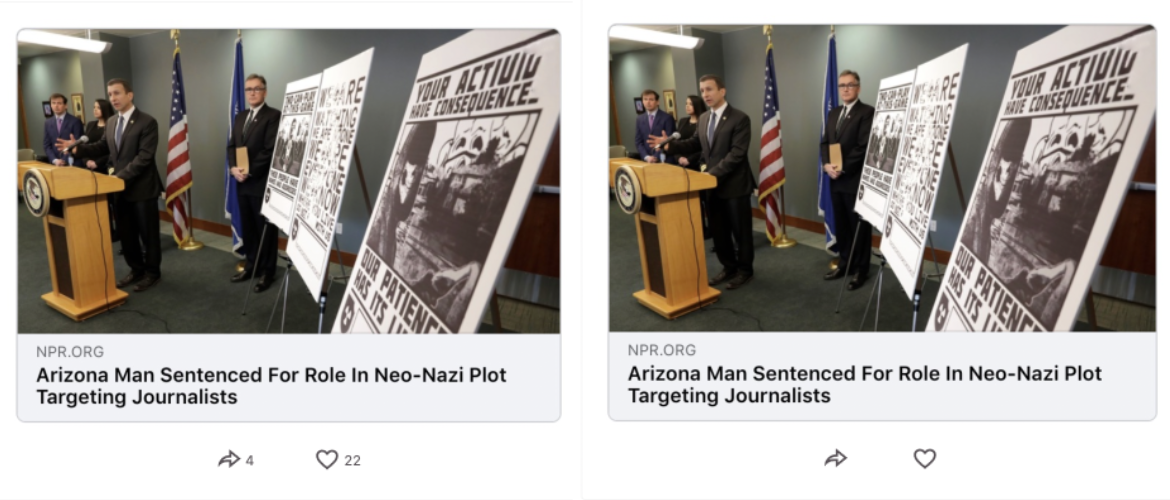}
    \caption{Example of a post shown in Yourfeed for treatment (left) and control (right) groups. Social cues for posts in the treatment condition were generated at random for each post for each participant.}
    \label{fig:stim}
\end{figure}

Participants were randomly assigned to one of two conditions. In the treatment condition, participants saw social cues  for each post in their feed (see Figure ~\ref{fig:stim}, left). In particular, each post contained a number of likes and number of shares. These numbers were simulated independently for each post for each participant as follows. For the number of likes, we followed prior work \cite{avram2020exposure} and sampled from a log-norm distribution (which matches the heavy-tailed distribution of engagement usually seen on social media \cite{vosoughi2018spread}), with mean 3 and standard deviation 0.8. For the number of shares, we divided the number of likes by a integer uniformly sampled between 5 and 20. This ensured that the like counts and share counts were somewhat, but not perfectly, related.  In the Control conditions, participants saw no social cues at all (see Figure ~\ref{fig:stim}, right).

\subsection{Pre-test rating task}
In addition to the main task described above, we also conducted a rating task to generate post-level features for each of the 276 posts used. We recruited N=1248 participants from the recruitment site Lucid to rate these posts, and filtered to N=872 who passed two attention checks. Participants where assigned to rate 40 randomly selected posts (of 276) on one of eight dimensions: 1) If you were to see the above article on social media, how likely would you be to share it?, 2) Are you familiar with the above headline (have you seen or heard about it before)?, 3) What is the likelihood that the above headline is true?, 4) Assuming the above headline is entirely accurate, how favorable would it be to Democrats versus Republicans?, 5) How provocative/sensational is this headline?, 6) How informative is this headline?, 7) How surprising is this headline?, and 8) How impactful is this headline? We average across participants to compute a single estimate for each dimension for each post (an average of 15.06 ratings per item per dimension). We then use these ratings below when assessing whether the effects of social cues differ based on these post features. 

\subsection{Post-level analysis}
For each post, we computed the proportion of times it was shared (total times shared / total times seen) across all participants within each condition. Separately for the control and treatment conditions, we fitted gradient boosting decision tree regressors \cite{ke2017lightgbm} to predict proportion share using the eight post features described above, and we used leave-one-out cross validation \cite{pedregosa2011scikit} to compute the squared error for each post. Larger errors indicate worse prediction, and thus greater unpredictability of how frequently a post would be shared, given the eight features. We also computed Shapley values \cite{lundberg2018consistent} to estimate feature importance and derive feature attributions for each post. We also re-ran the analysis to predict the proportion of times the post was liked.

\section{Results}
\subsection{Social Cues Increase Engagement}

\begin{figure}[h]
    \centering
    \includegraphics[width=0.99\textwidth]{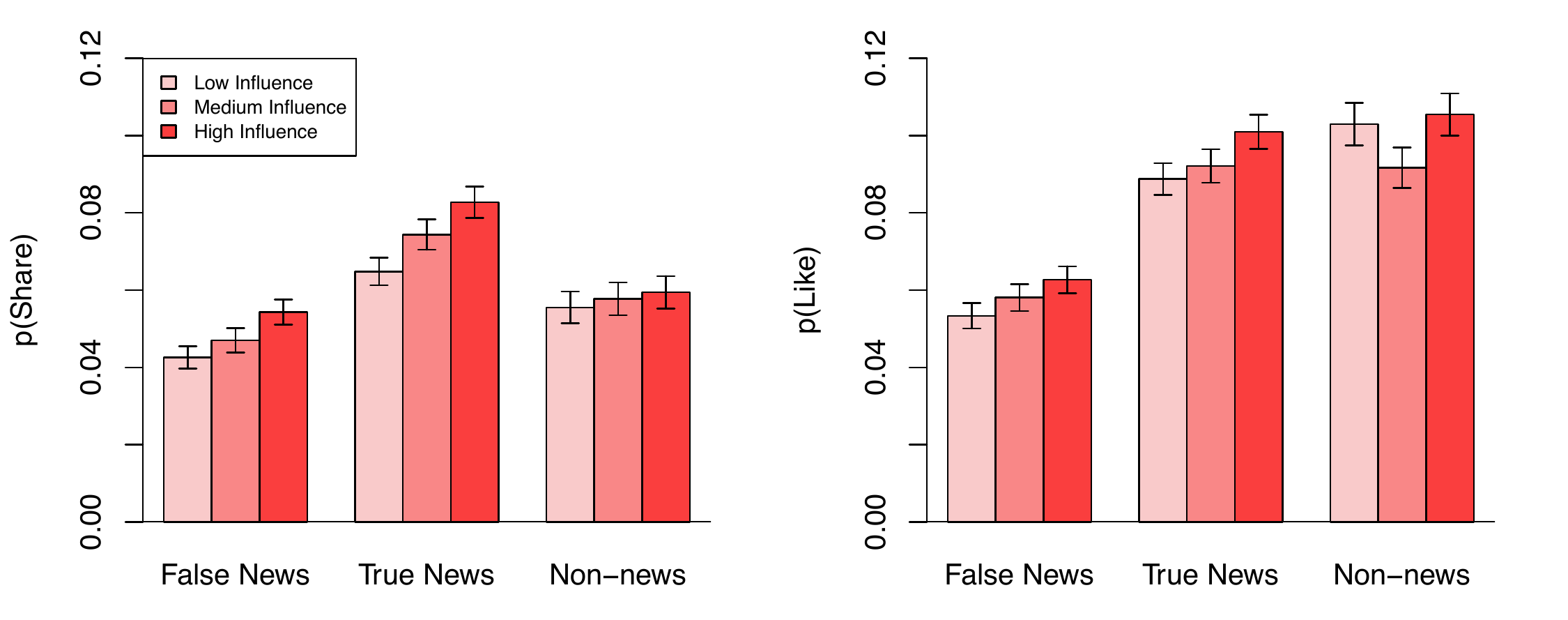}
    \caption{Sharing rates across three  social influence cues terciles in treatment data for false news, true news and non-news (e.g. opinion and mundane articles).}
    \label{fig:engo_rates}
    \end{figure}

To answer our first research question -- whether the degree of social influence increases the likelihood of engagement -- we examine the effect of the \textit{strength of the social cue} by considering only data from participants in the treatment condition. To quantify social cue strength, we log transform (1+) the number of shares a given post was reported as receiving, since they were generated from a log-normal distribution. 
We find that higher strength of displayed cues caused a higher likelihood to share posts ($b=0.007, p<0.001$, see Table \ref{table:rq1_main}). 
\begin{table}[ht]
\centering
\begin{tabular}{rrrrr}
  \hline
 & Estimate & Std. Error & z value & Pr($>$$|$z$|$) \\ 
  \hline
(Intercept) & 0.054 & 0.002 & 28.472 & 0.000 \\
  log(1 + btn\_shares) & 0.007 & 0.002 & 3.827 & 0.000 \\ 
   \hline
\end{tabular}
\caption{Regression predicting sharing for participants in the treatment condition, collapsed across post type.}
\label{table:rq1_main}
\end{table}

We also examined whether the effects of social cue strength on sharing differed across three types of posts (false news, true news, non-news posts). Although the effects of social cues on sharing were descriptively weaker for non-news posts (see Figure~\ref{fig:engo_rates}, left), the interactions between social cue strength and post type were not statistically significant ($p_{min}=0.241$, see Table \ref{tab:rq1_post}; likely due to fewer numbers of non-news posts and hence less statistical power to detect interaction effects). The effect of social cues for true news did not differ significantly from the effect for false news and non-news posts ($p=0.675$), and the effects on non-news posts also did not differ from news posts ($p=0.241$). A linear contrast of the two two-way interaction coefficients also found a non-significant difference ($p=0.208$), suggesting that the effects of social cues on the sharing of non-news posts and true news were similar. 

In addition to sharing posts on Yourfeed, participants also had the ability to like posts (see Figure ~\ref{fig:stim}). We also observe a similar trend for liking (see Figure~\ref{fig:engo_rates}, right). We find that higher social cue strength was associated with higher likelihood to like posts ($p=0.002$), and also that the interactions between social cue strength and post type were not statistically significant ($p_{min}=0.625$).   

\begin{table}[ht]
\centering
\begin{tabular}{rrrrr}
  \hline
 & Estimate & Std. Error & z value & Pr($>$$|$z$|$) \\ 
  \hline
(Intercept) & 0.041 & 0.003 & 14.838 & 0.000 \\ 
  log(1 + btn\_shares) & 0.008 & 0.003 & 2.856 & 0.004 \\ 
  true\_news & 0.024 & 0.004 & 5.575 & 0.000 \\ 
  nonnews & 0.014 & 0.005 & 2.963 & 0.003 \\ 
  log(1 + btn\_shares):true\_news & 0.002 & 0.004 & 0.420 & 0.675 \\ 
  log(1 + btn\_shares):nonnews & -0.005 & 0.004 & -1.173 & 0.241 \\ 
   \hline
\end{tabular}
\caption{Regression predicting sharing as a function of the interaction between engagement metric strength and headline type (false, true, non-news posts). Only participants in treatment condition were included in the analysis.}
\label{tab:rq1_post}
\end{table}

\begin{table}[ht]
\centering
\begin{tabular}{rlllp{0.15\linewidth}p{0.12\linewidth}}
  \hline
 & age & CRT &  partisanship & social media followers/following & social media usage \\ 
  \hline
  (Intercept) & \textbf{0.054}$^{***}$  & \textbf{0.052}$^{***}$  & \textbf{0.055}$^{***}$  & \textbf{0.055}$^{***}$  & \textbf{0.054}$^{***}$ \\ & (0.002) & (0.002) & (0.002) & (0.002) & (0.002) \\  veracity & \textbf{0.021}$^{***}$  & \textbf{0.02}$^{***}$  & \textbf{0.024}$^{***}$  & \textbf{0.023}$^{***}$  & \textbf{0.024}$^{***}$ \\ & (0.005) & (0.005) & (0.005) & (0.005) & (0.005) \\  log(1 + btnshares) & \textbf{0.008}$^{***}$  & \textbf{0.008}$^{**}$  & \textbf{0.008}$^{**}$  & \textbf{0.008}$^{**}$  & \textbf{0.008}$^{**}$ \\ & (0.002) & (0.002) & (0.002) & (0.002) & (0.002) \\  moderator & 0.005$^{.}$  & -0.004 & 0.003 & 0.001 & 0.002\\ & (0.002) & (0.002) & (0.002) & (0.002) & (0.002) \\  veracity:log(1 + btnshares) & 0.002 & 0.002 & 0.001 & 0.002 & 0.002\\ & (0.004) & (0.004) & (0.004) & (0.004) & (0.004) \\  veracity:moderator & 0.003 & 0.01$^{.}$  & \textbf{-0.019}$^{***}$  & -0.009$^{.}$  & 0.002\\ & (0.004) & (0.005) & (0.005) & (0.004) & (0.004) \\  log(1 + btnshares):moderator & -0.002 & -0.0 & -0.002 & -0.001 & 0.003\\ & (0.002) & (0.002) & (0.002) & (0.002) & (0.002) \\  veracity:log(1 + btnshares):moderator & 0.0 & -0.003 & 0.005 & 0.006 & 0.003\\ & (0.004) & (0.004) & (0.005) & (0.004) & (0.004)\\
   \hline
\end{tabular}
\caption{Regressions predicting sharing for treatment group only, with centered veracity, (-0.5=false, 0.5=true), z-scored item-level moderators and two-way clustered errors (post and participant). $^{.}$ refers to $p\leq0.1$, $^*$ and bolded refers to $p\leq.05$, $^{**}$ refers to $p\leq.01$, $^{***}$ refers to $p\leq.001$ with false-discovery rate adjusted p-values.}
\label{table:treatment_mod_id}
\end{table}

\begin{table}[h!]
\centering
\begin{tabular}{rllllllll}
  \hline
 & \small familiarity & \small truth & \small favorability & \small provocative & \small informative & \small surprising & \small impactful & news \\ 
  \hline
 (Intercept) & \textbf{0.054}$^{***}$  & \textbf{0.054}$^{***}$  & \textbf{0.054}$^{***}$  & \textbf{0.054}$^{***}$  & \textbf{0.054}$^{***}$  & \textbf{0.054}$^{***}$  & \textbf{0.054}$^{***}$  & \textbf{0.055}$^{***}$ \\ & (0.002) & (0.002) & (0.002) & (0.002) & (0.002) & (0.002) & (0.002) & (0.004) \\  log(1 + btnshares) & \textbf{0.007}$^{***}$  & \textbf{0.007}$^{***}$  & \textbf{0.007}$^{***}$  & \textbf{0.007}$^{***}$  & \textbf{0.007}$^{***}$  & \textbf{0.007}$^{***}$  & \textbf{0.007}$^{***}$  & 0.003\\ & (0.002) & (0.002) & (0.002) & (0.002) & (0.002) & (0.002) & (0.002) & (0.004) \\  moderator & \textbf{0.009}$^{***}$  & \textbf{0.01}$^{***}$  & -0.003 & 0.003 & \textbf{0.01}$^{***}$  & -0.003 & \textbf{0.008}$^{***}$  & -0.002\\ & (0.002) & (0.002) & (0.002) & (0.002) & (0.002) & (0.002) & (0.002) & (0.004) \\  log(1+btnshares):moderator & 0.001 & -0.001 & -0.004$^{.}$  & -0.003 & 0.001 & 0.001 & -0.001 & 0.006\\ & (0.002) & (0.002) & (0.002) & (0.002) & (0.002) & (0.002) & (0.002) & (0.004)\\
   \hline
\end{tabular}
\caption{Regressions predicting sharing for treatment group only, with centered treatment (-0.5=control, 0.5=treatment), z-scored item-level moderators and two-way clustered errors (post and participant). $^{.}$ refers to $p\leq0.1$, $^*$ and bolded refers to $p\leq.05$, $^{**}$ refers to $p\leq.01$, $^{***}$ refers to $p\leq.001$ with false-discovery rate adjusted p-values.}
\label{table:treatment_mod_post}
\end{table}

Overall, then, these results answer our first research question in the affirmative: seeing that a larger number of others have engaged with a post causes people to be more likely to engage with that post themselves. We also examined whether these findings were moderated by user- and post-level features. The effect of social cue strength on sharing was robust as it remained significant after controlling for five different user-level covariates (Table~\ref{table:treatment_mod_id}, row 3) and was not moderated by any user-level covariates (Table~\ref{table:treatment_mod_id}, row 7). Similarly, even though several post-level features (i.e. familiarity, truth, informative, impact) positively predicted sharing (Table~\ref{table:treatment_mod_post}, row 3), they did not moderate the effect of social cue strength on sharing (Table~\ref{table:treatment_mod_post}, row 4). Thus, the effects of social cue strength were similar not only for false news, true news, and non-news posts, but also across different demographic subgroups and posts with different features. 

\subsection{Social Cues Increase Sensitivity to Veracity}
\begin{figure}[h!]
    \centering
    \includegraphics[width=0.99\textwidth]{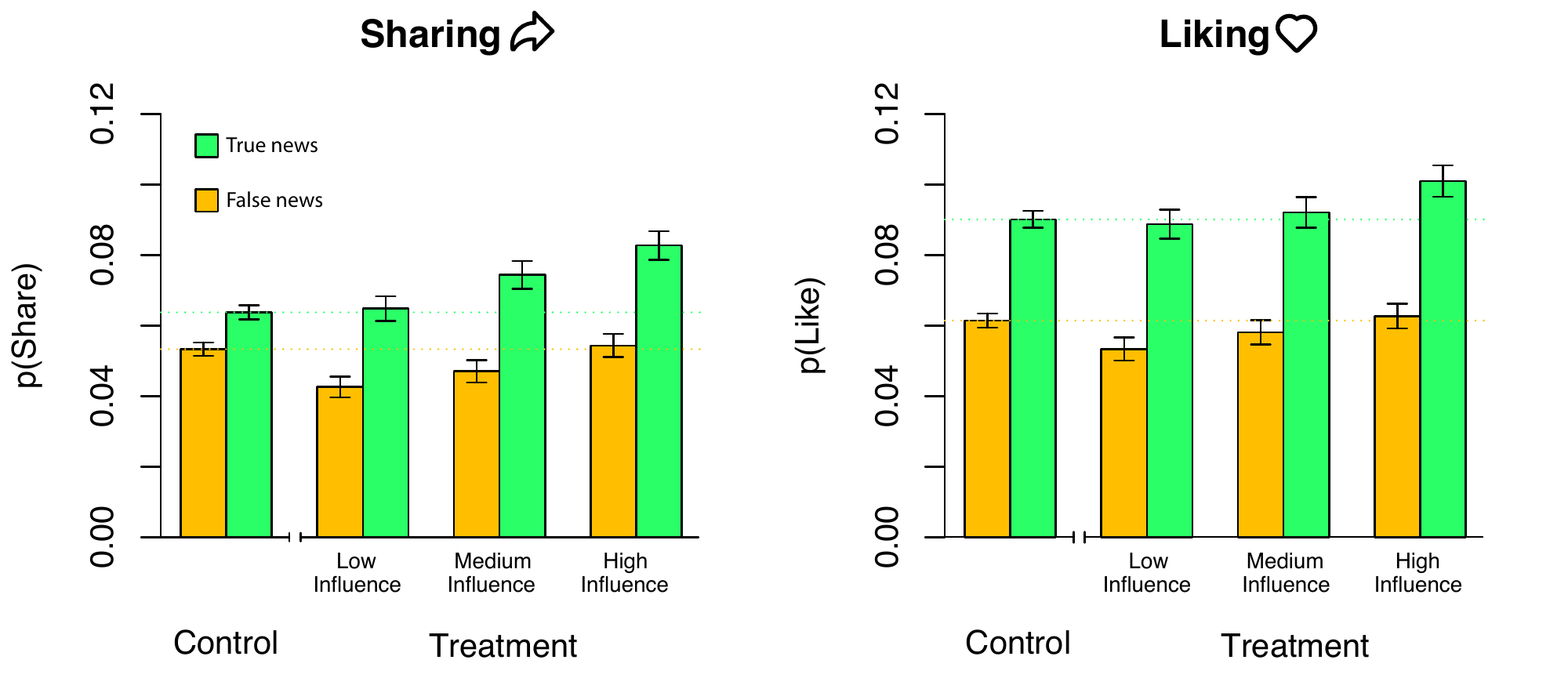}
    \caption{Left: Share rates for true and false news across conditions and engagement levels. Right: Like rates for true and false news across conditions and terciles of engagement levels.}
    \label{fig:engo_rates_veracity}
    \end{figure}

To answer our second research question -- whether adding social cues makes people less sensitive to headline veracity -- we compare sharing rates of true versus false news 
in treatment (where social cues were shown) with the control (where social  cues were not shown). 
As shown in Table \ref{tab:main_discernment}, overall, participants shared more true news than false news ($b=0.018, p<.001$), but the treatment group did not share more news than the control group ($b=0.002, p=0.230$). However, contrary to our expectations, we observed a significant \textit{increase} in sharing discernment in the treatment condition (interaction between headline veracity and condition, $b=0.015, p<0.001$; see Table \ref{tab:main_discernment}). That is, there was a larger difference in the likelihood of sharing true relative to false news in the treatment versus the control condition. Thus, the overall veracity of the news shared in the treatment condition was higher than in the control condition. In addition, as shown in Figure \ref{fig:engo_rates_veracity} (left), the difference in sharing of true and false content is larger at all levels of social influence for the treatment relative to control condition. We found that the pattern of results for liking are similar to sharing, albeit slightly weaker (see Figure \ref{fig:engo_rates_veracity} right). We observe an increase (though not significant) in liking discernment in treatment condition ($b=0.007, p=0.107$). 

The answer to our second research question, thus, is negative: social cues do not make people less sensitive to news veracity. In fact, it is the opposite: social cues increased sensitivity to veracity and increased engagement with true relative to false news.

\begin{table}[ht]
\centering
\begin{tabular}{rrrrr}

  \hline
 & Estimate & Std. Error & z value & Pr($>$$|$z$|$) \\ 
  \hline
(Intercept) & 0.060 & 0.001 & 59.959 & 0.000 \\ 
  veracity & 0.018 & 0.002 & 9.096 & 0.000 \\ 
  treatment & 0.002 & 0.002 & 1.201 & 0.230 \\ 
  veracity:treatment & 0.015 & 0.004 & 3.854 & 0.000 \\ 
   \hline
\end{tabular}
\caption{Regression table predicting sharing for control and treatment groups, with centered veracity (-0.5=false, 0.5=true) and centered treatment (-0.5=control, 0.5=treatment).}
\label{tab:main_discernment}
\end{table}


\begin{table}[ht]
\centering
\begin{tabular}{rlllp{0.15\linewidth}p{0.12\linewidth}}
  \hline
 & age & CRT &  partisanship & social media followers/following & social media usage \\ 
  \hline
(Intercept) & \textbf{0.06}$^{***}$  & \textbf{0.058}$^{***}$  & \textbf{0.061}$^{***}$  & \textbf{0.06}$^{***}$  & \textbf{0.06}$^{***}$ \\ & (0.001) & (0.001) & (0.001) & (0.001) & (0.001) \\  veracity & \textbf{0.016}$^{***}$  & \textbf{0.015}$^{***}$  & \textbf{0.017}$^{***}$  & \textbf{0.018}$^{***}$  & \textbf{0.018}$^{***}$ \\ & (0.002) & (0.002) & (0.002) & (0.002) & (0.002) \\  treatment & 0.002 & 0.003 & 0.002 & 0.003 & 0.002\\ & (0.002) & (0.002) & (0.002) & (0.002) & (0.002) \\  moderator & \textbf{0.003}$^{*}$  & \textbf{-0.005}$^{***}$  & \textbf{0.006}$^{***}$  & \textbf{0.005}$^{***}$  & \textbf{0.004}$^{***}$ \\ & (0.001) & (0.001) & (0.001) & (0.001) & (0.001) \\  veracity:treatment & \textbf{0.013}$^{**}$  & \textbf{0.012}$^{**}$  & \textbf{0.016}$^{***}$  & \textbf{0.014}$^{**}$  & \textbf{0.015}$^{**}$ \\ & (0.004) & (0.004) & (0.004) & (0.004) & (0.004) \\  veracity:moderator & \textbf{0.006}$^{**}$  & \textbf{0.012}$^{***}$  & \textbf{-0.017}$^{***}$  & \textbf{-0.008}$^{***}$  & 0.0\\ & (0.002) & (0.002) & (0.002) & (0.002) & (0.002) \\  treatment:moderator & 0.0 & 0.003 & \textbf{-0.009}$^{***}$  & \textbf{-0.01}$^{***}$  & 0.001\\ & (0.002) & (0.002) & (0.002) & (0.002) & (0.002) \\  veracity:treatment:moderator & -0.005 & -0.008$^{.}$  & 0.005 & 0.007 & 0.007$^{.}$ \\ & (0.004) & (0.004) & (0.005) & (0.004) & (0.004)\\ 
   \hline
\end{tabular}
\caption{Regressions predicting sharing across conditions, with centered veracity (-0.5=false, 0.5=true), centered treatment (-0.5=control, 0.5=treatment), z-scored item-level moderators and two-way clustered errors (post and participant). $^{.}$ refers to $p\leq0.1$, $^*$ and bolded refers to $p\leq.05$, $^{**}$ refers to $p\leq.01$, $^{***}$ refers to $p\leq.001$ with false-discovery rate adjusted p-values.}
\label{tab:overall_mod_id}
\end{table}

To provide additional evidence for this conclusion, we conducted user-level moderation analyses to investigate whether any covariates (five individual differences variables) moderated the two-way interaction between news veracity and condition. As shown in Table~\ref{tab:overall_mod_id} (row 4), participants with higher cognitive reflection test (CRT) scores shared fewer news posts overall, whereas those who are older, more Republican, have more social media followers/following, and use more social media shared more news posts overall. Participants who are older and have higher CRT scores were also more discerning, but Republicans and those with more followers/following were less discerning (Table~\ref{tab:overall_mod_id}, row 6). Interestingly, the effect of social cues on sharing was weaker for Republicans and those with more followers/following (Table~\ref{tab:overall_mod_id}, row 7). Crucially, the main effect of interest (two-way interaction between veracity and condition) remained significant after controlling for the five covariates and no three-way-interactions were significant, suggesting that the increased sensitivity to veracity in the treatment condition was robust across demographic subgroups. We also find the effect of condition on sharing was not significantly moderated by the eight different post-level features (Table~\ref{tab:overall_mod_post}).

\begin{table}[ht]
\centering
\begin{tabular}{rllllllll}
  \hline
 & \small familiarity & \small truth & \small favorability & \small provocative & \small informative & \small surprising & \small impactful & news \\ 
  \hline
 (Intercept) & \textbf{0.059}$^{***}$  & \textbf{0.059}$^{***}$  & \textbf{0.059}$^{***}$  & \textbf{0.059}$^{***}$  & \textbf{0.059}$^{***}$  & \textbf{0.059}$^{***}$  & \textbf{0.059}$^{***}$  & \textbf{0.059}$^{***}$ \\ & (0.001) & (0.001) & (0.001) & (0.001) & (0.001) & (0.001) & (0.001) & (0.001) \\  treatment & 0.003$^{.}$  & 0.003$^{.}$  & 0.003$^{.}$  & 0.003$^{.}$  & 0.003$^{.}$  & 0.003$^{.}$  & 0.003$^{.}$  & 0.003$^{.}$ \\ & (0.002) & (0.002) & (0.002) & (0.002) & (0.002) & (0.002) & (0.002) & (0.002) \\  moderator & \textbf{0.009}$^{***}$  & \textbf{0.007}$^{***}$  & \textbf{-0.006}$^{***}$  & \textbf{0.002}$^{*}$  & \textbf{0.009}$^{***}$  & -0.0 & \textbf{0.006}$^{***}$  & \textbf{0.002}$^{*}$ \\ & (0.001) & (0.001) & (0.001) & (0.001) & (0.001) & (0.001) & (0.001) & (0.001) \\  treatment:moderator & 0.003 & \textbf{0.004}$^{*}$  & -0.001 & -0.002 & 0.004$^{.}$  & -0.003$^{.}$  & 0.001 & -0.001\\ & (0.002) & (0.002) & (0.002) & (0.002) & (0.002) & (0.002) & (0.002) & (0.002)\\
   \hline
\end{tabular}
\caption{Regressions predicting sharing for across conditions with centered treatment (-0.5=control, 0.5=treatment), z-scored item-level moderators, and two-way clustered errors (post and participant). $^{.}$ refers to $p\leq0.1$, $^*$ and bolded refers to $p\leq.05$, $^{**}$ refers to $p\leq.01$, $^{***}$ refers to $p\leq.001$ with false-discovery rate adjusted p-values.}
\label{tab:overall_mod_post}
\end{table}

\subsection{Social Cues Increase Unpredictability of Success}
Having shown that participants engaged more with more popular posts (i.e. those that have been shared or liked by more people), we now turn to the question of whether the presence of social influence cues makes the ``success'' of any given post -- overall shares or likes -- more unpredictable. 

We trained gradient boosting decision trees to predict the overall sharing rate of each post using the eight post-level features obtained from the pre-test. We find that the mean squared error was significantly higher in the treatment than control condition ($p=0.007$), suggesting that the predictive accuracy of post sharing rate was worse in the treatment than control condition. Similarly, the mean squared error was also significantly higher in the treatment condition when the models were trained to predict the overall liking rate of each post ($p=0.007$). These results indicate that the presence of social cues (in the treatment condition) increased the unpredictability of engagement rate for any given post. In other words, in the presence of social influence, post features become less predictive of whether participants were going to engage with a given post and therefore it is more difficult to predict whether a post will be ``successful,'' given the features of that post.

By comparing the Shapley post feature attributions between the treatment and control conditions, we again find evidence consistent with the finding that social influence increases discernment. For the model trained to predict each post's sharing rate, the Shapley values for the truth feature was higher in the treatment than control condition ($p=0.012$), suggesting the veracity of a post had a stronger and more positive effect on sharing rate. We also find similar results for the model trained to predict each post's liking rate ($p=0.033$). These findings suggest that participants in the treatment condition engaged more with true than false news in part because they put more weight on veracity than participants in the control condition. 

These results answer our third research question: The presence of social cues increases the unpredictability of how frequently posts will be engaged with. 




\section{Discussion}
We find that when social cues were shown, users were more likely to engage with posts. This increase in engagement leads to inequality and unpredictability of success of posts. That is, the ``rich gets richer'' \cite{salganik2006experimental} because popular posts receive more engagement due simply to their popularity, while unpopular posts are less likely to be engaged with. Moreover, the presence of social cues makes the ``success'' of any given post -- i.e., likelihood of engagement -- more random or less predictable, which is also in line with \citet{salganik2006experimental}. Crucially, despite the increased in inequality and unpredictability, social cues did not reduce sharing discernment. If anything, contrary to our expectations, participants in the treatment condition (where social cues were shown) engaged with relatively more true than false news. Crucially, these results were not moderated by various user- or post-level features, underscoring the robustness of the findings.

Although we found similar patterns of results for sharing and liking, the effects were generally weaker for liking (though participants liked more than shared overall). It is likely because relative to sharing, liking is more passive, less consequential, and requires less engagement and attention \cite{scissors2016s}, and thus is less affected by changes in contextual cues. However, future work is needed to test this hypothesis.

Contrary to previous work that suggests that social cues drive the spread of misinformation \cite{avram2020exposure}, our results suggest that social cues can actually boost sharing discernment in certain circumstances. \citet{avram2020exposure} did not include a no-cue control, which means they were unable to compare the effects of cues to a baseline. What might explain why, within the treatment condition, they found a differential effect of social cues on sharing for true versus false news, while we did not? A critical difference between our design and theirs is that their game was specifically framed as a fact-checking exercise - and on every headline, they were prompted as to whether they would fact-check that claim. This constitutes a heavy accuracy prime, which has been shown to substantially change sharing intentions \cite{pennycook2021shifting, epste, epstein2021social}. Thus, we expect that the pattern we observed (absent the accuracy prime) will be more reflective of actual behavior online than the pattern they observed.

Our results have important implications for the design of social media. Our work suggests that social cues might increase engagement and attentiveness, which could explain why participants in the treatment condition were more discerning in terms of what they shared and liked. But at the same time, too much engagement can drive huge inequalities and unpredictability in the system \cite{salganik2006experimental}, which have their own downsides. Thus, platform designers much navigate this design space in service of an information environment that benefits its users accordingly. Instagram has explored removing the number of post likes, but ultimately found little effect \cite{hide_like}. Another related consideration is downvotes and how it affects engagement and discernment. For example, recently, YouTube removed the number of downvotes for a video, but kept the number of upvotes, and the effects on engagement and discernment remain unclear. Future work could investigate the effects of the downvote button, and if it could make people even more engaged/discerning. 

Our work also highlights a possible scalable design intervention for mitigating misinformation. Rather than removing low-quality content entirely (which ignites free speech issues) or annotate them with a warning flag (which could induce the implied truth effect \citep{pennycook2020implied}), platforms could artificially deflate the social cues for low-quality content. This possibility is exciting but also has its own challenges (e.g. people may stop believing in the cues themselves) that future work should explore and address.

Our study has several limitations that pave the way for future work. It is important to note that we used simulated social cues that were random across posts and participants. Using simulated values allows us to dissociate the effects of social cue strength and post features, but it raises two two potential avenues for future work. First, the simulated distributions of social cues we used may not capture the absolute magnitude of posts in the wild. Future work should explore people's perceptions of the distribution of social cues because these distributions could affect engagement. Second, people have expectations about the relative amount of likes and shares certain posts will get, and when those expectations are broken (e.g. a very odd post with a lot of shares/likes or a very interesting post with very few), it might cause them to question the validity of the context and thus engage differently. Future work could follow \citet{salganik2006experimental} and \citet{epstein2021social} to surface actual social cues from other participants interacting with Yourfeed. 

Another limitation of our study is that sharing decisions were hypothetical, and did not correspond to sharing a given post with anyone in particular. It has been shown that hypothetical sharing in a survey context correlates with actual sharing on Twitter at the headline level \cite{mosleh2020self}, but it is possible that some of the psychological tradeoffs, social pressures, and consequences involved with sharing a post with your followers are absent in Yourfeed. Future work could tie sharing on Yourfeed to actual real-world outcomes. 

Finally, our sample and results are constrained to a US setting. It is plausible that the nature and effect of social cues varies across cultures, a possibility which in turn poses large challenges for platforms attempting to design a ``universal'' platform that spans cultural contexts. As such, future work should investigate how these effects vary across different cultural contexts \cite{arechar2022understanding}. 

\bibliographystyle{ACM-Reference-Format}
\bibliography{sample-base}

\end{document}